\documentclass[%
 aip,
 amsmath,amssymb,
 reprint,%
]{revtex4-1}

\usepackage{graphicx}
\usepackage{dcolumn}
\usepackage{bm}

\usepackage[utf8]{inputenc}
\usepackage[T1]{fontenc}
\usepackage{mathptmx}
\usepackage{etoolbox}

\makeatletter
\def\@email#1#2{%
 \endgroup
 \patchcmd{\titleblock@produce}
  {\frontmatter@RRAPformat}
  {\frontmatter@RRAPformat{\produce@RRAP{*#1\href{mailto:#2}{#2}}}\frontmatter@RRAPformat}
  {}{}
}%
\makeatother
\begin{document}

\preprint{AIP/123-QED}

\title[]{Optical Response in Spintronic Poisson Bolometers}

\author{Ziyi Yang\textsuperscript{\dag}} 
\affiliation{These authors contributed equally to this work}

\affiliation{Elmore Family School of Electrical and Computer Engineering, Purdue University, West Lafayette, Indiana 47907, USA}%

\author{Sakshi Gupta\textsuperscript{\dag}}
\affiliation{These authors contributed equally to this work}
\affiliation{Department of Physics and Astronomy, Purdue University, West Lafayette, Indiana 47907, USA}%

\author{Jehan Shalabi\textsuperscript{\dag}}

\affiliation{These authors contributed equally to this work}
\affiliation{Elmore Family School of Electrical and Computer Engineering, Purdue University, West Lafayette, Indiana 47907, USA}%

\author{Leif Bauer}
\affiliation{Elmore Family School of Electrical and Computer Engineering, Purdue University, West Lafayette, Indiana 47907, USA}%

\author{Daien He}
\affiliation{Elmore Family School of Electrical and Computer Engineering, Purdue University, West Lafayette, Indiana 47907, USA}%

\author{Mohamed A. Mousa}
\affiliation{Elmore Family School of Electrical and Computer Engineering, Purdue University, West Lafayette, Indiana 47907, USA}%

\author{Angshuman Deka}
\affiliation{Birck Nanotechnology Center, Purdue University, West Lafayette, Indiana 47907, USA}

\author{Zubin Jacob*}
\affiliation{Elmore Family School of Electrical and Computer Engineering, Purdue University, West Lafayette, Indiana 47907, USA}%
\affiliation{Department of Physics and Astronomy, Purdue University, West Lafayette, Indiana 47907, USA}%

\email{zjacob@purdue.edu}

\date{\today}

\begin{abstract}

Analog bolometers based on temperature-dependent phase-transition materials such as vanadium oxide (VOx) and barium titanate (BTO) represent the state of the art in uncooled infrared detectors. Recently, the first room-temperature spintronic Poisson bolometer based on magnetic tunnel junctions (MTJs) was proposed and demonstrated as a promising infrared detector. Unlike conventional bolometers, the spintronic Poisson bolometer operates in a probabilistic regime dominated by Poissonian noise, where the response is governed by resistance fluctuations arising from thermally activated magnetization transitions. Spontaneous transitions between two metastable magnetic states occur even in the absence of incident light, and the transition probability increases under illumination. In this work, we experimentally study the statistical properties of the optical response of the spintronic Poisson bolometer under illumination. We demonstrate that transitions in spintronic Poisson bolometers, both in the absence and presence of light, exhibit Poissonian behavior, with transition rates and interarrival times modulated by incident radiation. Under illumination, we observe a 153\% increase in the count rate accompanied by a 70\% reduction in interarrival time. These results establish spintronic Poisson bolometers as a promising platform for high-speed, and high-sensitivity infrared detection at room temperature.

\end{abstract}

\maketitle

\section{Introduction}

\begin{figure*}[ht!]
\includegraphics[width= 1\textwidth]{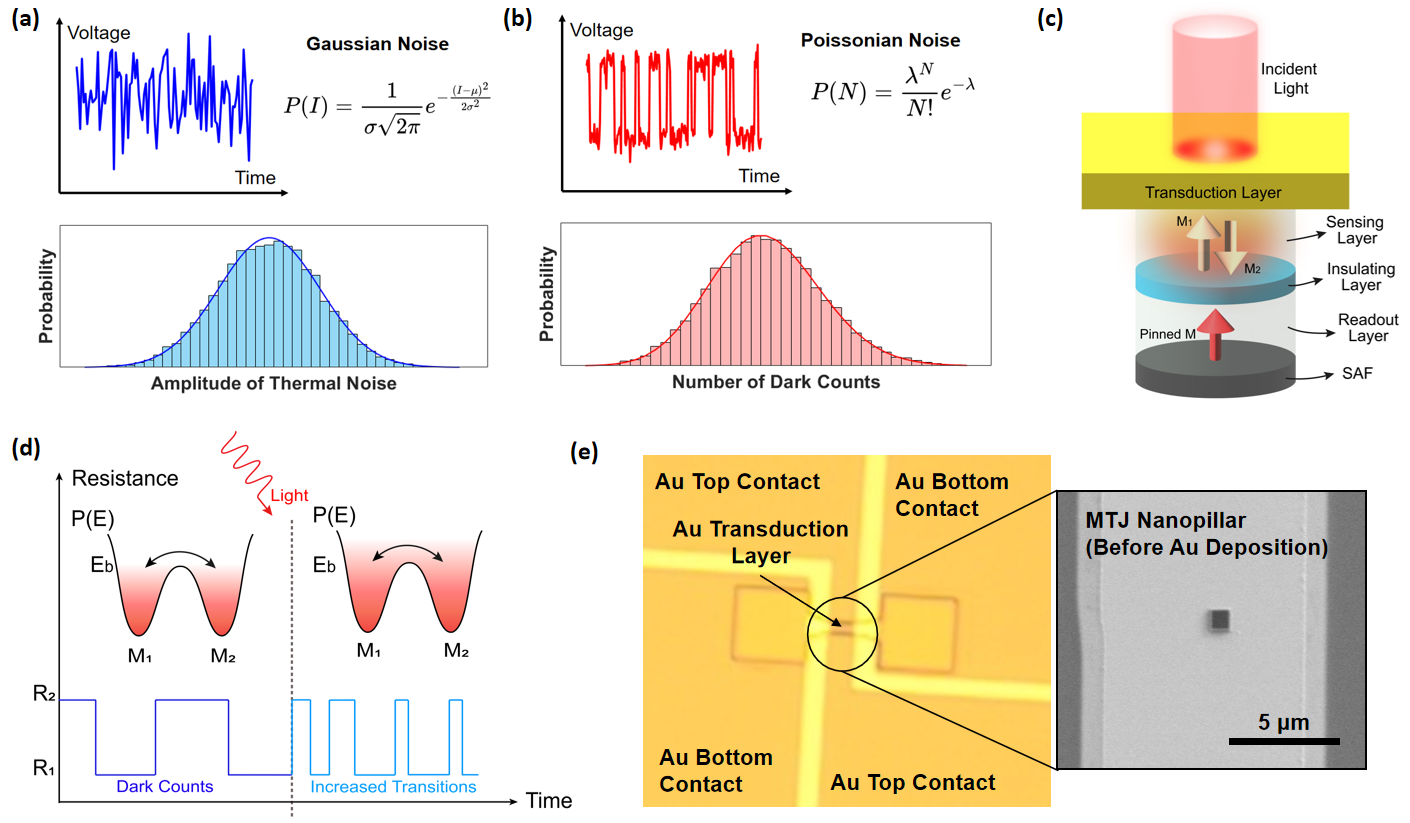}
\caption{\label{fig:one} (a) In analog detectors, thermal noise and Johnson noise follow a Gaussian distribution as they stem from thermal agitation of a large number of electrons. (b) In digital detectors, dark counts follow a Poisson distribution as they stem from discrete independent events happening at a constant mean rate, such as thermal excitations and quantum tunneling effects in single photon detectors (SPDs), or magnetization flips governed by exponentially distributed interarrival times in spintronic Poisson bolometers (SPB). (c) Device structure. The main structure has five layers: a transduction layer, a magnetic free layer, a tunnel barrier, a magnetic fixed layer, and a synthetic antiferromagnetic layer (SAF). The magnetization M of the readout layer is pinned by the SAF. The magnetization of the sensing layer flips between M1 and M2 probabilistically due to heat. (d)  When no light is incident, natural heat in the device causes discrete transitions between these two magnetization directions, which we refer to as dark counts. When light is incident, a hot spot is formed in the transduction layer, which thermalizes through a magnetic tunnel junction (MTJ). This increases the probability of transition in the MTJs sensing layer, leading to an increased count rate, which is read out through the device’s resistance. (e) Left: Optical image of the device. Right: SEM image of the MTJ nanopillar before Au deposition.}
\end{figure*}

Bolometers are thermal detectors in which incident light causes a temperature-induced resistance change that is read out electrically \cite{YADAV2022113611}. Infrared bolometers are essential for emerging applications in thermal imaging, autonomous navigation, astronomy, and time-of-flight sensing \cite{Bao2023, Lahiri2012, Sivaprakasam2020, Kaushal2017, anderson2023advancements, rodriguez2023remote, PhysRevX.14.041005}. State-of-the-art uncooled bolometers are analog detectors based on temperature-dependent phase transition of materials such as vanadium oxide (VOx) and barium titanate (BTO) \cite{yeh2020performance, scott2022sensing}. Other uncooled bolometers are based on semiconductor and metallic materials with temperature-dependent resistance, including amorphous silicon and platinum\cite{doi:10.1021/acsphotonics.9b01198,doi:10.1021/acs.nanolett.1c02972, doi:10.1021/acs.nanolett.3c03076}. In VOx bolometers, major noise sources are Johnson noise, thermal noise, generation–recombination, and $1/f$ noise \cite{philipose2023analyzing, 10.1063/5.0049633, du2013graphene}. Similarly, the dominant noise in barium titanate (BTO) bolometers and graphene bolometers is Johnson noise \cite{Hanel:61, du2013graphene, PhysRevB.96.165431}. All of the uncooled bolometers aforementioned are analog detectors that produce a continuous voltage or current proportional to the incoming signal, and are dominated by Gaussian noise, as shown in Fig.~\ref{fig:one}(a). 

Recently, the first uncooled spintronic Poisson bolometer, called the spintronic ultrafast nanoscale bolometer (SUN bolometer, or SUN-B) \cite{leif_nanolett} was proposed. Unlike analog bolometers, the spintronic Poisson bolometers convert incoming signals into discrete counting events and are dominated by Poissonian noise, as shown in Fig.~\ref{fig:one}(b). In spintronic Poisson bolometers, magnetic tunnel junctions (MTJs) are engineered with an energy barrier of the order of thermal infrared energy at room temperature. It exploits switching between two discrete magnetization states by intrinsic thermal fluctuations and nanoscale heating induced by incident radiation. These digital probabilistic transitions in the absence of light are referred to as dark counts, and when under illumination, are bright counts. It was demonstrated that the count rate increases due to the incident light, validating the device’s response through the Néel-Arrhenius law \cite{PhysRevB.104.094433}. The spintronic Poisson bolometer achieved a noise equivalent temperature difference (NETD) of 103 mK at a 25 Hz frame rate, and high readout speeds of 10 MHz, which is faster than commercial vanadium dioxide detectors. The high sensitivity and the high readout speed make it a promising candidate for state-of-the-art infrared detection.

In other studies in the field, stochastic MTJs have been utilized for random number generation, probabilistic computing, and neuromorphic computing \cite{choi2014magnetic, joshi2020mtj, zhang2021recent, maciel2019magnetic, zink2022review}. The stochastic magnetization transitions, known as magnetic random telegraph noise (RTN), exhibit relaxation times that follow an exponential distribution, consistent with Poisson event statistics \cite{hayakawa2021nanosecond,2021relaxation}. Similarly, detection-based studies have shown exponentially distributed dark-count interarrival times \cite{leif_nanolett}. Notably, previous studies have examined the statistics of MTJ transitions only in the absence of incident light.

In this work, we extend the study to the optical response of the spintronic Poisson bolometer under incident radiation. The spintronic Poisson bolometer operates in a noise-dominated probabilistic regime, where the response is governed by random resistance fluctuations arising from thermally activated magnetization switching. We demonstrate that both dark-count and bright-count statistics in spintronic Poisson bolometers are well described by the Poisson distribution, with the mean count rate modulated by incident radiation. Changes in the mean number of counts indicate enhanced thermal switching between the two magnetization states in the MTJ sensing layer, serving as a key to quantifying incident radiation. The spintronic Poisson bolometer thus offers a fundamentally different detection paradigm based on an indirect, thermally mediated light–matter interaction, leading to Poisson statistics that are independent of the light source statistics.

The Poissonian nature of these counts enables time-resolved measurements with high fidelity and superior signal-to-noise ratio (SNR) compared to analog systems under identical conditions. The spintronic Poisson bolometer also exhibits distinct statistics compared to other digital detectors due to the absence of after-pulsing \cite{WANG2019202} and its intrinsic Poisson response, irrespective of light statistics. We also investigate the device's non-probabilistic response. Excitation with a pulsed laser produces current transients, providing insights into heat transfer and transport within the device. This systematic experimental investigation of the response statistics of spintronic Poisson bolometers under illumination advances our understanding of the underlying device physics and guides optimization of sensitivity, response time, dynamic range, and spectral response. Overall, we establish the spintronic Poisson bolometer as a promising platform for probabilistic, high-speed, and high-sensitivity infrared detection at room temperature.

\section{Results and Discussion}

 The spintronic Poisson bolometer studied here consists of five layers: a transduction layer at the top, a magnetic sensing (free) layer, a nonmagnetic insulating tunnel barrier, a magnetic readout (fixed) layer, and a synthetic antiferromagnetic (SAF) layer, as shown in Fig.~\ref{fig:one}(c). The latter four layers form a magnetic tunnel junction (MTJ) that exhibits tunnel magnetoresistance. The magnetization of the readout layer is pinned by the underlying SAF layer, which stabilizes an out-of-plane (perpendicular) magnetization in the fixed layer. The sensing layer possesses two stable magnetization states (M1 and M2). Due to the presence of spin-polarized current in the device, the resistance depends on the sensing layer's magnetic orientation. 
 
 Even without illumination, thermal fluctuations induce spontaneous transitions between the two sensing-layer states, producing dark counts. Illumination generates a hot spot in the transduction layer, which thermalizes through the MTJ stack and increases the probability of transitions in the sensing layer. This results in an increased count rate observed through MTJ resistance readout (Fig.~\ref{fig:one}(d)). The behavior of these transitions follows the Néel–Arrhenius law for thermally activated switching\cite{2021relaxation}:
 \begin{equation}
 \label{eq:N-A law}
\tau_{M_1,(M_2)} = \tau_0 \exp\left( \frac{E_b}{k_B T} \right)
\end{equation}

\noindent where \(\tau_0\) is the attempt time, \(k_B\) is Boltzmann’s constant, $T$ is the temperature, and $E_b$ is the energy barrier:

 \begin{equation}
E_b = \frac{M V (H_k \mp H)}{2}
\end{equation}

\noindent where $M$ is the magnetization, $H_k$ is the magnetic anisotropy field, V is the volume of the sensing layer, and H is the magnetic field present in the sensing layer. Fig.~\ref{fig:one}(e) shows an optical image of the fabricated spintronic Poisson bolometer device and an SEM image of the MTJ nanopillar before Au deposition. Detailed MTJ layer composition and dimensions are shown in section S1 of the supplementary material. The device shows a measured tunnel magnetoresistance (TMR) ratio of 1.5\%, and detailed TMR results can be found in Figure S2 of the supplementary material.

We next present a general model for detector signal and response time. For analog bolometers such as VOx, the detector signal is an output voltage, whereas for the spintronic Poisson bolometer, the signal is the count rate. The relationship between incident power and detector signal is quantified through the temperature increase of the bolometer, which depends on the thermal properties:

\begin{equation}
\Delta T = \frac{\varepsilon P / A}{\left(G_{\mathrm{th}}^{2} + \omega^{2} C_{\mathrm{th}}^{2}\right)^{1/2}}
\end{equation}

\noindent where $\Delta T$ is the temperature change of the detector, $\varepsilon$ is emissivity, $P$ is the incident radiant power, $A$ is the effective detector area, $G_{\mathrm{th}}$ is the thermal conductance, $C_{\mathrm{th}}$ is the thermal capacitance, and $\omega$ is the angular modulation frequency \cite{8845784}.

The temperature-to-signal conversion is characterized by:

\begin{equation}
 k = \frac{dV}{dT} 
\end{equation}

\noindent which reflects the sensitivity of the detector output to temperature change.\cite{8845784} For VOx, $k$ is constant over the operating range, whereas for the spintronic Poisson bolometer, the change in count rate follows the Néel-Arrhenius relation \cite{leif_nanolett}. For small changes in temperature, the count rate increase is

\begin{equation}
 \label{eq:N-A law}
dN = \frac{2}{\tau_0} \exp( \frac{-E_b}{k_B T})*\frac{E_b}{k_B T^2}*dT
\end{equation}

\noindent where $dN$ is the increase in counts due to a temperature rise $dT$, $\tau_0$ is the relaxation time, $E_b$ is the energy barrier, $k_B$ is the Boltzmann constant (detailed derivation is included in the supplementary material section S2). Thus, in both conventional analog bolometers and digital spintronic Poisson bolometers, the signal is linearly dependent on temperature.

The detector response time is given by:
\begin{equation}
\tau_{total} =\tau_{diffusion}+\tau_{thermal}+\tau_{intrinsic} 
\end{equation}

\noindent where $\tau_{\text{diffusion}}$ is the diffusion time, $\tau_{\text{thermal}}$ is the thermal time constant, and $\tau_{\text{intrinsic}}$ represents intrinsic system dynamics as magnetic spin switching. For VOx bolometers, the response time is dominated by thermal time constant (10 ms), whereas for the spintronic Poisson bolometer, it is governed by $\tau_{intrinsic}$, i.e., the speed of magnetic spin switching (2.2-3.5 ns) \cite{leif_nanolett}. Because the spintronic Poisson bolometer is a probabilistic Poisson detector, deterministic readout of a transition is further limited by approximately 4.6 times the relaxation time (0.1-10 $\mu s$) (detailed derivation is included in the supplementary material section S3). Overall, the spintronic Poisson bolometer is faster than $VO_x$ by three orders of magnitude, making it a strong candidate for high-speed detector applications.

To study the spintronic Poisson bolometer response to incident light over a range of pulse widths, powers, and wavelengths, we used three different laser sources: (1) a continuous-wave (CW) 405 nm laser, (2) an 808 nm CW laser with power modulation by an arbitrary waveform generator, and (3) a 405 nm picosecond laser with a 20 ps pulse width. A schematic of the corresponding power-time profiles is shown in Fig.~\ref{fig:two} (a).

\subsection{\label{sec:level2}Device Response to 808 nm Modulated Laser}

\begin{figure*}[ht!]
\includegraphics[width=\textwidth]{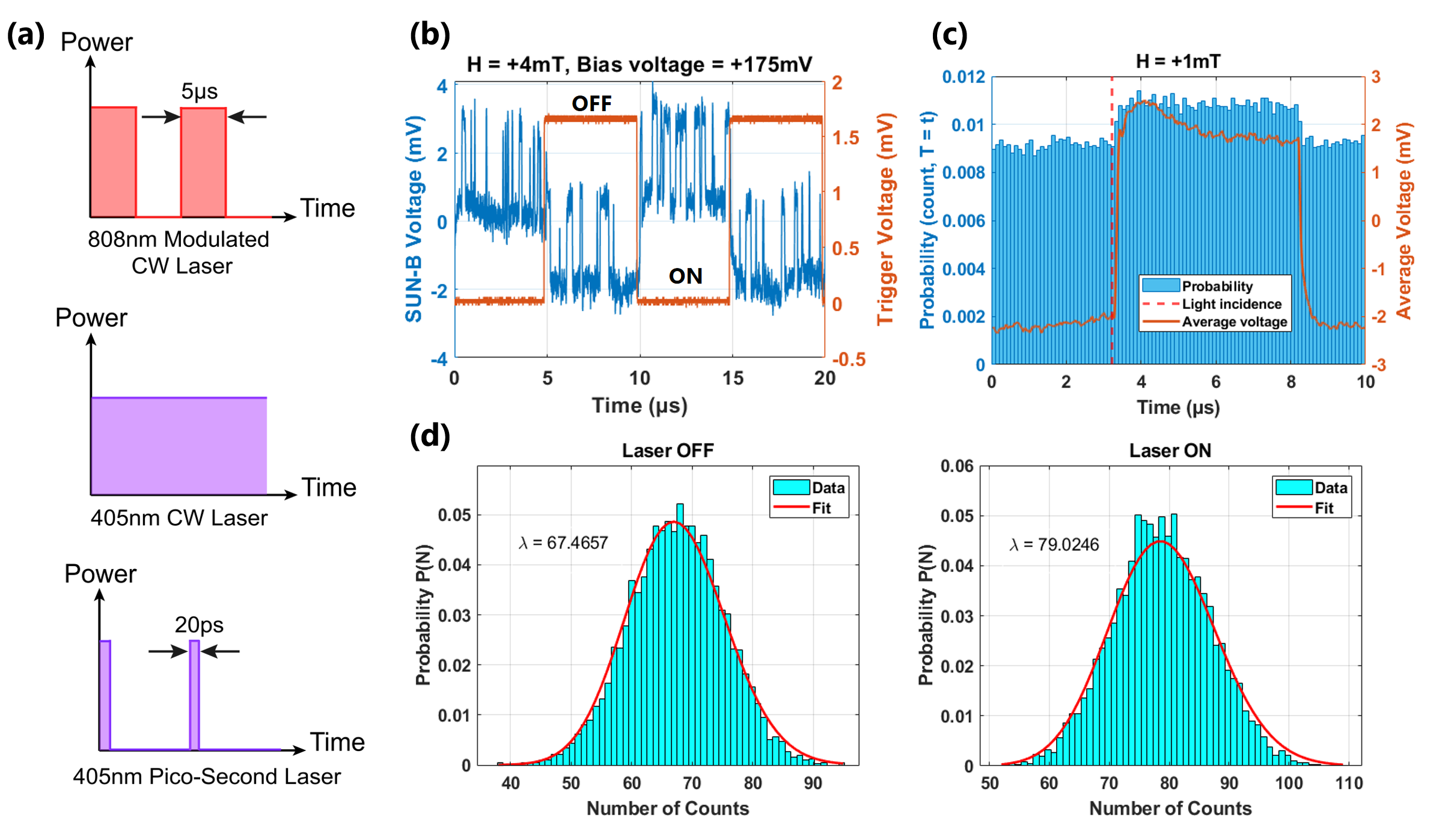}
\caption{\label{fig:two} (a)Schematic of the power-time relationship of 3 laser sources. We study the device response to 3 types of laser sources. The first laser is a continuous-wave (CW) 405 nm laser. The second laser is a modulated 808 nm CW laser. The power output of this laser is modulated by an arbitrary waveform generator (AWG). The third laser is a 405 nm picosecond laser that has a pulse width as small as 20 ps. (b) Raw readout signal. The orange curve is the laser trigger signal. (c) Time correlation. We divide each period into 100 bins (100 ns per bin) and calculate the probability of a count happening in each bin. The red dashed line marks the time of the trigger of one of the periods when it turns the laser on, which occurs at t = 3233.6 ns. The system has around 80 ns latency between the laser trigger and light incidence on the device. We can see a stabilized count probability increase at t = 3366.7 ns, which is around 53 ns after the light is incident when considering the latency. (d) Count statistics. We divide the data into 20 $\mu s$ time bins and calculate the number of counts happening in each time bin. For both laser-on and off regions, the number of counts follows a Poisson distribution. With the fitted mean number of counts $\lambda$, we can calculate the mean count rates for laser off and on regions, which are 3.373285 Mcps and 3.95123 Mcps, respectively. This shows around a 17\% increase in the count rate due to incident light.
 }
\end{figure*}

We first investigate the spintronic Poisson bolometer's response to an 808 nm near-infrared continuous (CW) laser whose intensity is modulated by an arbitrary waveform generator (AWG). This modulation produces periodic on-off illumination, enabling a direct comparison of count statistics in the presence and absence of light. The stochasticity of this device occurs within a magnetic-field window from $-2$ mT to $+4$ mT (see Figure S3 of the supplementary material for measurements of dark count rate versus the bias field). We choose a bias field of +4 mT to operate away from the maximum count-rate condition, making the count-rate increase induced by laser illumination more distinguishable while avoiding potential count saturation. Due to the relatively small TMR ratio, we use a bias voltage of 175 mV to ensure that the parallel and antiparallel states are clearly resolvable while avoiding excessively high bias voltages that could cause degradation or dielectric breakdown of the MgO tunnel barrier due to electrostatic discharge.

The incident average optical power is 25 mW, and the laser spot has a 3 $\mu m$ full-width half-maximum (FWHM), corresponding to an intensity of 3.5 mW/$\mu m^2$. The modulation frequency is 100 kHz with a 50\% duty cycle, producing alternating 5 $\mu s$ bright and dark intervals. We perform the measurement using a bias-tee and low-noise amplifier (LNA) in the setup and read out the AC part of the signal with an oscilloscope.

Figure ~\ref{fig:two} (b) shows the device response to the 808 nm modulated laser. It can be seen that stochastic counts are superimposed on the non-probabilistic voltage signal. When the laser is ON, we observe a decrease in the device resistance due to laser heating of the tunnel barrier as well as an increase in count rate. To understand the statistics of ON and OFF regions, we divide each period into 100 bins (100 ns per bin), and the probability of detecting a count in each bin is computed over 6000 periods, as shown in Fig.~\ref{fig:two} (c). The red dashed line marks the time of the trigger of one of the periods when it turns the laser on, and the orange line is the voltage of the device averaged over 6000 periods. The probability of a count remains constant over both ON/OFF regions, indicating stochastic behavior which is characteristic of Poisson statistics. When the laser is turned ON, we observe that the probability of a count increases 53 ns after the light is incident. In addition, we noticed a transient decay of the average voltage response, which occurs at the first 2.5 $\mu s$ after light is incident, likely governed by thermal equilibration.  A detailed discussion of this transient decay can be found in section S7 of the supplementary material. The experimental setup and characterization of system latency can be found in section S8 of the supplementary material.

We then separate the data into laser-ON and laser-OFF regions and study the probability distribution of the number of counts \(P(N)\) for each region. We divide the data into 100 time bins and calculate the number of counts that occur in each time bin as shown in Fig.~\ref{fig:two} (d). We observe that for both laser OFF and laser ON cases, the count distribution follows a Poisson distribution \cite{van1992stochastic}:

\begin{equation} 
P(N) = \frac{\lambda ^ N}{N!}e^{-\lambda}, N = 0,1,2,...,\infty.
\end{equation}

From the Poisson fit of the probability distribution in Fig.~\ref{fig:two} (d), the mean count rates for laser off and on regions are calculated to be 3.3 Mcps and 3.9 Mcps, respectively. This shows around a 17\% increase in the count rate due to incident light. The shifted mean number of counts manifests increased temperature and energy in the sensing layer, and thus increased probability of transitions according to equation \ref{eq:N-A law}. These results suggest that the spintronic Poisson bolometer demonstrates Poissonian statistics for both dark counts and bright counts with incident radiation.



\begin{figure*}[t!]
    \centering
    \includegraphics[width=1\textwidth]{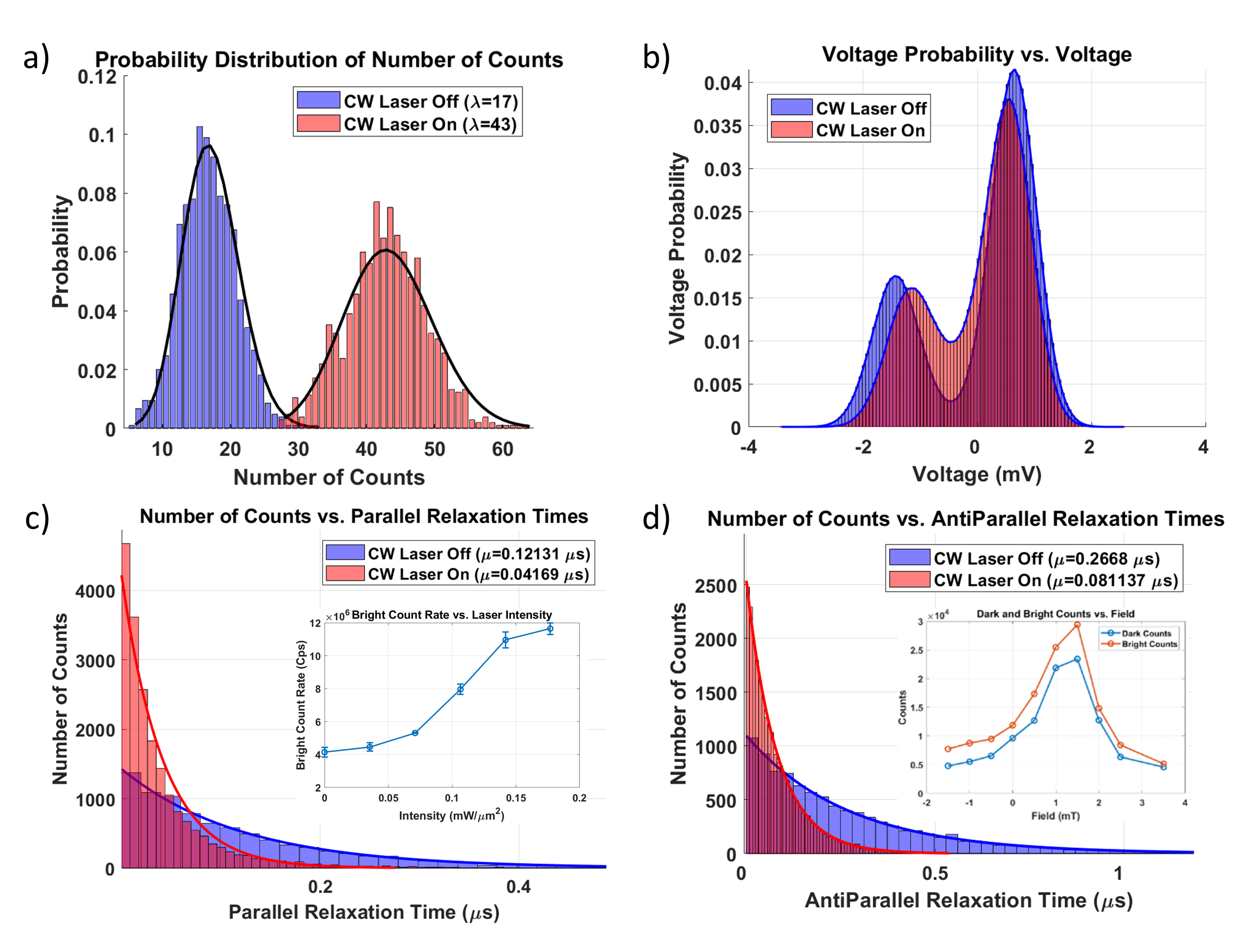}
    \caption{(a) Probability histogram of detector counts over \(3.8~\mu\text{s}\) intervals under 405 nm CW laser illumination. The count statistics follow a Poisson distribution, as shown by the black fit curve. When the laser is off, the mean count is \(\lambda\)= 17; under illumination, the mean increases to \(\lambda\) = 43, indicating light-induced switching events. (b) Voltage state distributions under dark and illuminated conditions show two distinct peaks corresponding to the parallel (low-resistance) and antiparallel (high-resistance) magnetization states. (c) Histogram of relaxation times spent in the parallel (low-voltage) state before transitioning to the antiparallel (high-voltage) state. The distribution follows an exponential decay, consistent with Poissonian statistics. Illumination reduces the mean relaxation time by 66\%, increasing the switching rate. Inset: Bright count rate versus laser intensity shows a significant increase in counts in response to higher optical power. (d) Histogram of relaxation times in the antiparallel state before switching to the parallel state. Like the parallel case, the distribution follows an exponential profile, with mean relaxation time decreasing by 70\% under illumination. Inset: Field dependence of both dark and bright count rates shows a maximum near H=1 mT, corresponding to the field strength where the energy barrier between magnetic states is minimized and switching is most probable.}
    \label{fig:Rot_Curve}
\end{figure*}

\begin{figure*}[t!]
    \includegraphics[width=1\textwidth]{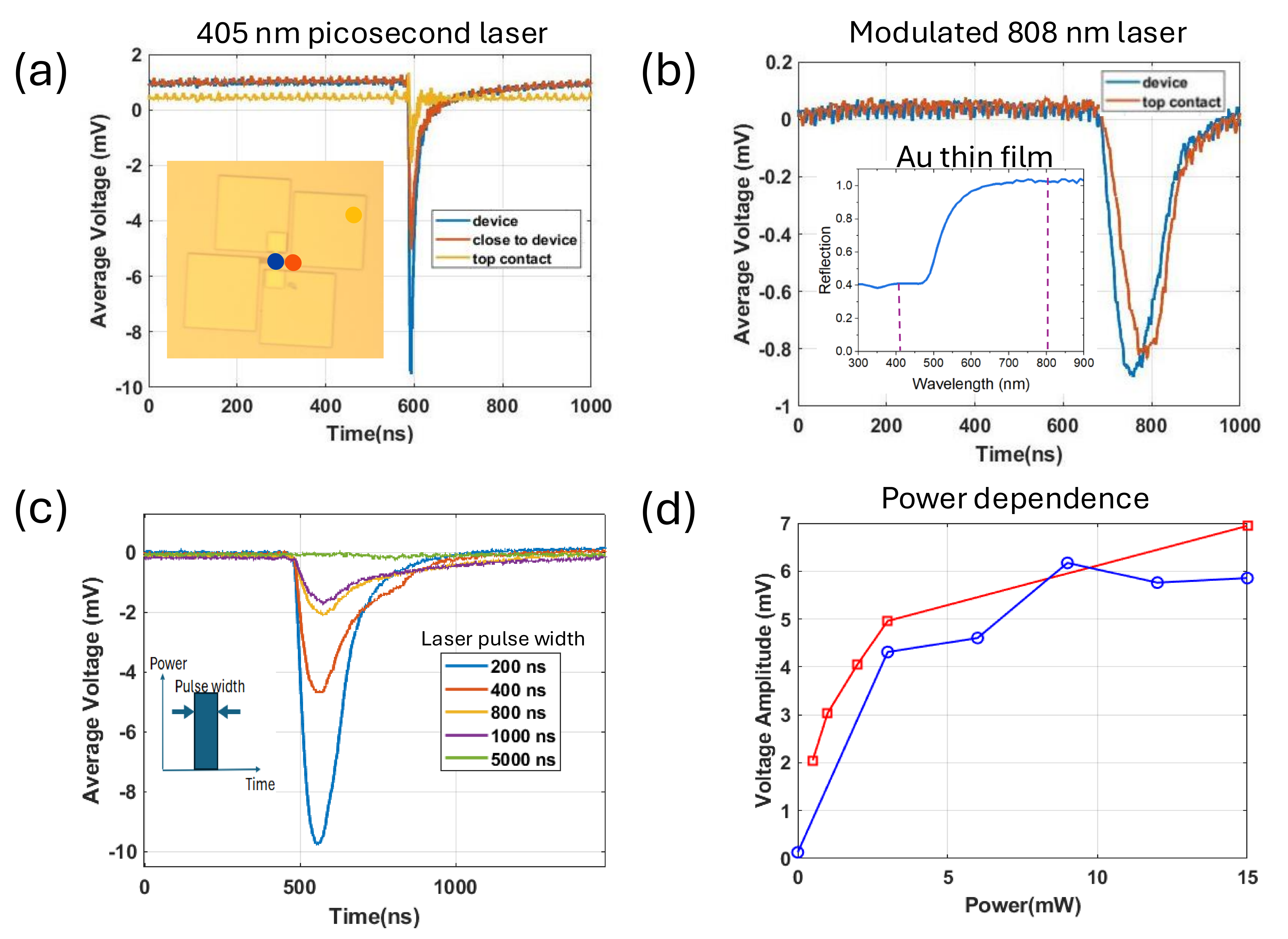}
    \caption{(a) For a 405 nm picosecond laser, the laser-induced response is observed in the spintronic Poisson bolometer readout when light is incident on or near the device. (b) We observe different behavior with the 808 nm laser, where the photocurrent response does not reduce further away from the device. Inset is the reflection spectrum of the Au top contact. (c)The pulse width dependence is studied by varying the duty cycle of the 808 nm modulated laser. We observe that for larger pulse widths, the photocurrent is reduced, which signifies power dependence. (d) The peak power dependence of the laser-induced voltage is linear at low powers and saturates at high powers. The red curve corresponds to the 405 nm laser, and the blue curve corresponds to the 808 nm laser.}
    \label{fig:Rot_Curve}
\end{figure*}

\subsection{\label{sec:level2}Device Poissonian Response to 405 nm CW Laser}

In this section, we examine the spintronic Poisson bolometer’s response to a 405 nm CW laser of intensity of $0.18 \mathrm{ mW}/\mu\mathrm{m}^2$ on the device and analyze the statistical behavior of both dark and bright counts, assessing the count statistics. The device registers a dark count rate of ~4.5 Mcps. When illuminated with a 405 nm CW laser, the count rate increases to ~11.4 Mcps—a 153 \% increase. The gold top contact of the device absorbs approximately \(60\%\) of incident light at 405 nm, compared to \(1\%\) absorption at 808 nm. Thus, there is a higher count rate increase under a 405nm laser illumination than under 808nm, indicating a strong wavelength dependence of the statistical response.

Figure 3(a) presents histograms of detector counts over \(3.8~\mu\text{s}\) intervals under dark and illuminated conditions. Both distributions follow Poissonian fits, with mean number of counts shifting from \(\lambda\) = 17 (dark) to \(\lambda\) = 43 (bright). This demonstrates that incident photons increase the probability of stochastic switching in the magnetic tunnel junction (MTJ), consistent with thermally activated transitions. Figure 3(b) shows the voltage state probability distributions for both conditions. Each case exhibits two distinct peaks corresponding to the parallel (low resistance) and antiparallel (high resistance) magnetization states of the sensing layer. Under illumination, a voltage shift is observed as a result of changes in magnetoresistance, providing further confirmation of light-induced modulation.

To further investigate the switching dynamics, we analyze the relaxation time, i.e., the time between consecutive magnetization transitions. Figures 3(c) and 3(d) present histograms of relaxation times for the parallel and antiparallel states under both laser-off and laser-on conditions. In both cases, the distributions exhibit exponential decay, consistent with a Poisson process, as described by: 

\begin{equation}
P(\tau) = \frac{1}{\mu} e^{-\frac{1}{\mu}\tau}
\end{equation} 

\noindent where \(\mu\) is the mean relaxation time. Under illumination, \(\mu\) decreases significantly: from \(121  \text{ns}\) to \(41 \text{ns}\) in the parallel state, and from \(266  \text{ ns}\) to \(81 \text{ns}\) in the antiparallel state, resulting in an increased switching rate. This confirms that optical excitation enhances the probability of transition events, and is consistent with Néel–Arrhenius' law (see Eq. 1), where the switching rate increases as the energy barrier decreases via optical heating that increases the temperature of the sensing layer.

The inset of Fig. 3(c) demonstrates that as the laser intensity increased from \(0\) to \(0.18\,\frac{\mathrm{mW}}{\mu\mathrm{m}^2}\), the count rate nearly triples, from  4.5 Mcps to 11.4 Mcps—approaching saturation at higher intensities. Since the detector output is the count rate, the responsivity is proportional to the local slope dN/dP, where N is the number of counts and P is the incident power. In the approximately linear regime, dN/dP remains nearly constant, leading to stable responsivity and reliable power reconstruction. At higher incident powers, the count rate begins to saturate and dN/dP decreases, reducing the detector responsivity and limiting the usable dynamic range. At very low powers, the slope is also reduced due to the weak count-rate response. Therefore, optimal detector responsivity is achieved within the intermediate linear regime. For the specific device and measurement shown in Figure 3(c) inset, the operating regime for optimal responsivity, i.e., the linear response regime, is roughly 0.07 to 0.14 $\text{mW/}\mu \text{m}^2$ illumination intensity, identified from the approximately linear region of the count-rate curve. We expect that the dynamic range can be improved by engineering the energy barrier or by actively cooling the device to a lower operating temperature \cite{leif_nanolett}.

The inset of Fig. 3(d) shows that both dark and bright count rates follow a Gaussian dependence on the applied magnetic field, peaking near H = 1 mT where we operate our device. At this field bias, the energy barrier is minimum, leading to an increased transition rate. These results confirm the field-tunable nature of Poissonian distributed thermal switching events in the spintronic Poisson bolometer under 405 nm illumination, with optical modulation of both count rate and relaxation dynamics.

\subsection{\label{sec:level2}Laser Induced Current in MTJ-based detector due to a pulsed picosecond laser}

Finally, we investigate the response of the spintronic Poisson bolometer to nanosecond and picosecond pulsed lasers. When the laser pulse is incident on the device, the absorbed light produces phonons and hot electrons. The top contact of the device is made up of 75 nm Au, which is known to induce bulk photocurrent due to pulsed laser. Since the device has an electrical readout, this photocurrent response is read out in addition to the MTJ tunnel magneto-resistance. We studied the induced-current response due to a 405 nm picosecond laser with a pulse width of 20 ps (Figure 4a) and an 808 nm modulated laser with a nanosecond (200-1000 ns) pulse width (Figure 4b) by analyzing the average voltage of the response. With high power of incident radiation, we observe a photocurrent in the metallic Au top contact. 

The positional dependence of the photocurrent is studied for the two laser signals. For the 405 nm ps laser, we observe that the current is highest when light is incident on top of the device, and it reduces significantly when light is incident away from the device. This implies the presence of hot electron effects. Additionally, it takes the same time to get to the peak current for different powers of light, while the relaxation time depends on the power of incident radiation. For 808nm, we observe that the current induced is largely position-independent, indicating the presence of hot electron effects. The effect of the laser on the top and bottom contacts was also studied, with no major differences found in voltage, indicating the absence of thermo-voltaic effects. The presence of hot-electron effects at 405 nm can be explained by differences in the reflection spectrum. For 405 nm, Au is significantly absorptive (reflection=40\%) while at 808 nm Au is highly reflective (reflection=99\%). 

The induced current is studied for different pulse widths (Figure 4c). For smaller pulse widths, since the peak power of the laser is higher, the induced current is higher. The induced-current duration is shorter than an 800 ns pulse width, as it is a non-equilibrium effect. The power dependence is also analyzed (Figure 4d). The induced current is found to be linearly proportional to low incident powers and saturates at higher powers. Thus, we demonstrate a non-probabilistic response with ultrafast pulsed light and show a response dominated by photocurrent and hot-electron effects.

\section{Conclusion}

In conclusion, we have experimentally demonstrated that spintronic Poisson bolometers exhibit Poissonian response statistics under both dark and illuminated conditions, independent of the statistics of the incident light source. Time-resolved measurements confirm that both dark and bright counts follow Poisson distributions, with interarrival times exhibiting exponential behavior. This validates the indirect thermally activated switching mechanism governed by the Néel-Arrhenius model. These findings establish spintronic Poisson bolometers as a new class of digital probabilistic infrared detectors with high-speed, high-sensitivity, room-temperature operation. By leveraging their inherent Poisson statistics, spintronic Poisson bolometers present a promising platform for next-generation time-resolved infrared detection and effectively address the limitations of current state-of-the-art analog detectors.

\section{Supplementary Material}

The supplementary material provides detailed MTJ layer composition and dimensions. It also includes a detailed derivation of equation (5) and the integration time required for a spintronic Poisson bolometer to produce a deterministic response. Additional device characterization results, such as TMR measurements, field dependence of dark count rate, and non-probabilistic device response, are included. The transient decay in the average voltage response and system latency is also discussed.

\begin{acknowledgments}
This work is partially supported by the Elmore Chaired Professorship of Purdue University.
\end{acknowledgments}

\section*{Data Availability Statement}
Data available on request from the authors.

\bibliography{refs}

@article{YADAV2022113611,
title = {Advancements of uncooled infrared microbolometer materials: A review},
journal = {Sensors and Actuators A: Physical},
volume = {342},
pages = {113611},
year = {2022},
issn = {0924-4247},
doi = {https://doi.org/10.1016/j.sna.2022.113611},
url = {https://www.sciencedirect.com/science/article/pii/S0924424722002497},
author = {P.V. Karthik Yadav and Isha Yadav and B. Ajitha and Abraham Rajasekar and Sudha Gupta and Y. {Ashok Kumar Reddy}}
}

@article{PhysRevB.96.165431,
  title = {Dynamically tunable extraordinary light absorption in monolayer graphene},
  author = {Safaei, Alireza and Chandra, Sayan and V\'azquez-Guardado, Abraham and Calderon, Jean and Franklin, Daniel and Tetard, Laurene and Zhai, Lei and Leuenberger, Michael N. and Chanda, Debashis},
  journal = {Phys. Rev. B},
  volume = {96},
  issue = {16},
  pages = {165431},
  numpages = {10},
  year = {2017},
  month = {Oct},
  publisher = {American Physical Society},
  doi = {10.1103/PhysRevB.96.165431},
  url = {https://link.aps.org/doi/10.1103/PhysRevB.96.165431}
}

@book{van1992stochastic,
  title={Stochastic processes in physics and chemistry},
  author={Van Kampen, Nicolaas Godfried},
  volume={1},
  year={1992},
  publisher={Elsevier}
}

@article{10.1063/5.0049633,
    author = {Ye, Ming and Zha, Jiajia and Tan, Chaoliang and Crozier, Kenneth B.},
    title = {Graphene-based mid-infrared photodetectors using metamaterials and related concepts},
    journal = {Applied Physics Reviews},
    volume = {8},
    number = {3},
    pages = {031303},
    year = {2021},
    month = {07},
    issn = {1931-9401},
    doi = {10.1063/5.0049633},
    url = {https://doi.org/10.1063/5.0049633}

}

@article{doi:10.1021/acs.nanolett.3c03076,
author = {Lien, Max R. and Wang, Nan and Guadagnini, Silvia and Wu, Jiangbin and Soibel, Alexander and Gunapala, Sarath D. and Wang, Han and Povinelli, Michelle L.},
title = {Black Phosphorus Molybdenum Disulfide Midwave Infrared Photodiodes with Broadband Absorption-Increasing Metasurfaces},
journal = {Nano Letters},
volume = {23},
number = {21},
pages = {9980-9987},
year = {2023},
doi = {10.1021/acs.nanolett.3c03076},
    note ={PMID: 37883580},

URL = { 
    
        https://doi.org/10.1021/acs.nanolett.3c03076
    
    

},
eprint = { 
    
        https://doi.org/10.1021/acs.nanolett.3c03076
    
    

}

}

@ARTICLE{8845784,
  author={Kazmierkowski, Marian P.},
  journal={IEEE Industrial Electronics Magazine}, 
  title={Infrared and Terahertz Detectors, Third Edition [Book News]}, 
  year={2019},
  volume={13},
  number={3},
  pages={53-54},
  doi={10.1109/MIE.2019.2929349}}

@article{PhysRevB.104.094433,
  title = {Magnetic relaxation time for an ensemble of nanoparticles with randomly aligned easy axes: A simple expression},
  author = {Chalifour, Artek R. and Davidson, Jonathon C. and Anderson, Nicholas R. and Crawford, Thomas M. and Livesey, Karen L.},
  journal = {Phys. Rev. B},
  volume = {104},
  issue = {9},
  pages = {094433},
  numpages = {11},
  year = {2021},
  month = {Sep},
  publisher = {American Physical Society},
  doi = {10.1103/PhysRevB.104.094433},
  url = {https://link.aps.org/doi/10.1103/PhysRevB.104.094433}
}

@article{PhysRevX.14.041005,
  title = {A 25-micrometer Single-Photon-Sensitive Kinetic Inductance Detector},
  author = {Day, Peter K. and Cothard, Nicholas F. and Albert, Christopher and Foote, Logan and Kane, Elijah and Eom, Byeong H. and Basu Thakur, Ritoban and Janssen, Reinier M. J. and Beyer, Andrew and Echternach, Pierre M. and van Berkel, Sven and Hailey-Dunsheath, Steven and Stevenson, Thomas R. and Dabironezare, Shahab and Baselmans, Jochem J. A. and Glenn, Jason and Bradford, C. Matt and Leduc, Henry G.},
  journal = {Phys. Rev. X},
  volume = {14},
  issue = {4},
  pages = {041005},
  numpages = {20},
  year = {2024},
  month = {Oct},
  publisher = {American Physical Society},
  doi = {10.1103/PhysRevX.14.041005},
  url = {https://link.aps.org/doi/10.1103/PhysRevX.14.041005}
}

@article{WANG2019202,
title = {Afterpulsing effects in SPAD-based photon-counting communication system},
journal = {Optics Communications},
volume = {443},
pages = {202-210},
year = {2019},
issn = {0030-4018},
doi = {https://doi.org/10.1016/j.optcom.2019.03.039},
url = {https://www.sciencedirect.com/science/article/pii/S0030401819302342},
author = {Chen Wang and Jingyuan Wang and Zhiyong Xu and Jianhua Li and Rong Wang and Jiyong Zhao and Yimei Wei},
}

@article{doi:10.1021/acs.nanolett.1c02972,
author = {Chen, Chen and Li, Cheng and Min, Seunghwan and Guo, Qiushi and Xia, Zhenyang and Liu, Dong and Ma, Zhenqiang and Xia, Fengnian},
title = {Ultrafast Silicon Nanomembrane Microbolometer for Long-Wavelength Infrared Light Detection},
journal = {Nano Letters},
volume = {21},
number = {19},
pages = {8385-8392},
year = {2021},
doi = {10.1021/acs.nanolett.1c02972},
    note ={PMID: 34606292}
}

@article{doi:10.1021/acsphotonics.9b01198,
author = {Wu, Yangbo and Qu, Zhibo and Osman, Ahmed and Cao, Wei and Khokhar, Ali Z. and Soler Penades, Jordi and Muskens, Otto L. and Mashanovich, Goran Z. and Nedeljkovic, Milos},
title = {Mid-Infrared Nanometallic Antenna Assisted Silicon Waveguide Based Bolometers},
journal = {ACS Photonics},
volume = {6},
number = {12},
pages = {3253-3260},
year = {2019},
doi = {10.1021/acsphotonics.9b01198} 

}

@inproceedings{anderson2023advancements,
  title={Advancements in linear-mode photon-counting HgCdTe e-APDs for space and defense applications},
  author={Anderson, P Duke and McCurdy, James and Huebner, Adam and Sawatsky, Austin and Schaake, Christopher and Cook, Grady and Sun, Xiaoli and Mitra, Pradip},
  booktitle={Advanced Photon Counting Techniques XVII},
  volume={12512},
  pages={31--36},
  year={2023},
  organization={SPIE}
}

@article{rodriguez2023remote,
  title={Remote exploration and monitoring of geothermal sources: A novel method for foliar element mapping using hyperspectral (VNIR-SWIR) remote sensing},
  author={Rodriguez-Gomez, Cecilia and Kereszturi, Gabor and Jeyakumar, Paramsothy and Pullanagari, Reddy and Reeves, Robert and Rae, Andrew and Procter, Jonathan N},
  journal={Geothermics},
  volume={111},
  pages={102716},
  year={2023},
  publisher={Elsevier}
}

@article{yeh2020performance,
  title={Performance improvement of Y-doped VOx microbolometers with nanomesh antireflection layer},
  author={Yeh, Tsung-Han and Tsai, Cheng-Kang and Chu, Shao-Yu and Lee, Hsin-Ying and Lee, Ching-Ting},
  journal={Optics express},
  volume={28},
  number={5},
  pages={6433--6442},
  year={2020},
  publisher={Optical Society of America}
}

@article{scott2022sensing,
  title={Sensing performance of sub-100-nm vanadium oxide films for room temperature thermal detection applications},
  author={Scott, Ethan A and Singh, Manish K and Barber, John P and Rost, Christina M and Ivanov, Sergei and Watt, John and Pete, Douglas and Sharma, Peter and Lu, Tzu-Ming and Harris, C Thomas},
  journal={Applied Physics Letters},
  volume={121},
  number={20},
  year={2022},
  publisher={AIP Publishing}
}

@inproceedings{choi2014magnetic,
  title={A magnetic tunnel junction based true random number generator with conditional perturb and real-time output probability tracking},
  author={Choi, Won Ho and Lv, Yang and Kim, Jongyeon and Deshpande, Abhishek and Kang, Gyuseong and Wang, Jian-Ping and Kim, Chris H},
  booktitle={2014 IEEE International Electron Devices Meeting},
  pages={12--5},
  year={2014},
  organization={IEEE}
}

@article{joshi2020mtj,
  title={From MTJ device to hybrid CMOS/MTJ circuits: A review},
  author={Joshi, Vinod Kumar and Barla, Prashanth and Bhat, Somashekara and Kaushik, Brajesh Kumar},
  journal={IEEE Access},
  volume={8},
  pages={194105--194146},
  year={2020},
  publisher={IEEE}
}

@article{zhang2021recent,
  title={Recent progress and challenges in magnetic tunnel junctions with 2D materials for spintronic applications},
  author={Zhang, Lishu and Zhou, Jun and Li, Hui and Shen, Lei and Feng, Yuan Ping},
  journal={Applied Physics Reviews},
  volume={8},
  number={2},
  year={2021},
  publisher={AIP Publishing}
}

@article{maciel2019magnetic,
  title={Magnetic tunnel junction applications},
  author={Maciel, Nilson and Marques, Elaine and Naviner, L{\'\i}rida and Zhou, Yongliang and Cai, Hao},
  journal={Sensors},
  volume={20},
  number={1},
  pages={121},
  year={2019},
  publisher={MDPI}
}

@article{zink2022review,
  title={Review of magnetic tunnel junctions for stochastic computing},
  author={Zink, Brandon R and Lv, Yang and Wang, Jian-Ping},
  journal={IEEE Journal on Exploratory Solid-State Computational Devices and Circuits},
  volume={8},
  number={2},
  pages={173--184},
  year={2022},
  publisher={IEEE}
}

@article{philipose2023analyzing,
  title={Analyzing the bolometric performance of vanadium oxide thin films modified by carbon nanotube dispersions},
  author={Philipose, Usha and Littler, Chris and Jiang, Yan and Naciri, Alia and Harcrow, Michael and Syllaios, AJ},
  journal={Materials},
  volume={16},
  number={4},
  pages={1534},
  year={2023},
  publisher={MDPI}
}

@article{Hanel:61,
author = {Rudolf A. Hanel},
journal = {J. Opt. Soc. Am.},
number = {2},
pages = {220--224},
publisher = {Optica Publishing Group},
title = {Dielectric Bolometer: A New Type of Thermal Radiation Detector},
volume = {51},
month = {Feb},
year = {1961},
url = {https://opg.optica.org/abstract.cfm?URI=josa-51-2-220},
doi = {10.1364/JOSA.51.000220},

}

@article{du2013graphene,
  title={Graphene-based bolometers},
  author={Du, Xu and Prober, Daniel E and Vora, Heli and Mckitterick, Chris},
  journal={arXiv preprint arXiv:1308.4065},
  year={2013}
}

@article{leif_nanolett,
  title={Exploiting Spintronics at Room Temperature for Long-Wave Infrared Nanophotonic Digital Bolometers},
  author={Bauer, Leif and Deka, Angshuman and Mousa, Mohamed A and Gupta, Sakshi and He, Daien and Huang, Sijay and Prasad, Bhagwati and Santos, Tiffany and Ray, Biswajit and Jacob, Zubin},
  journal={Nano Letters},
  year={2025},
  publisher={ACS Publications}  

}

@article{Bao2023,
  
   author = {Fanglin Bao and Xueji Wang and Shree Hari Sureshbabu and Gautam Sreekumar and Liping Yang and Vaneet Aggarwal and Vishnu N. Boddeti and Zubin Jacob},
   doi = {10.1038/s41586-023-06174-6},
   issn = {14764687},
   issue = {7971},
   journal = {Nature},
   title = {Heat-assisted detection and ranging},
   volume = {619},
   year = {2023}
}

@misc{Lahiri2012,
  
   author = {B. B. Lahiri and S. Bagavathiappan and T. Jayakumar and John Philip},
   doi = {10.1016/j.infrared.2012.03.007},
   issn = {13504495},
   issue = {4},
   journal = {Infrared Physics and Technology},
   title = {Medical applications of infrared thermography: A review},
   volume = {55},
   year = {2012}
}

@inproceedings{Sivaprakasam2020,
   
   author = {Vasanthi Sivaprakasam and Heath E. Gemar and Michael K. Yetzbacher and Abbie T. Watnik},
   doi = {10.1364/oe.477499},
   issn = {21622701},
   booktitle = {Optics InfoBase Conference Papers},
   title = {Multi-Spectral SWIR Lidar for imaging and spectral discrimination through partial obscurations},
   year = {2020}
}

@misc{Kaushal2017,

   author = {Hemani Kaushal and Georges Kaddoum},
   doi = {10.1109/COMST.2016.2603518},
   issn = {1553877X},
   issue = {1},
   journal = {IEEE Communications Surveys and Tutorials},
   title = {Optical Communication in Space: Challenges and Mitigation Techniques},
   volume = {19},
   year = {2017}
}

@article{hayakawa2021nanosecond,
  title={Nanosecond random telegraph noise in in-plane magnetic tunnel junctions},
  author={Hayakawa, Keisuke and Kanai, Shun and Funatsu, Takuya and Igarashi, Junta and Jinnai, Butsurin and Borders, WA and Ohno, H and Fukami, S},
  journal={Physical review letters},
  volume={126},
  number={11},
  pages={117202},
  year={2021},
  publisher={APS}
}

@article{2021relaxation,
  title = {Theory of relaxation time of stochastic nanomagnets},
  author = {Kanai, Shun and Hayakawa, Keisuke and Ohno, Hideo and Fukami, Shunsuke},
  journal = {Phys. Rev. B},
  volume = {103},
  issue = {9},
  pages = {094423},
  numpages = {12},
  year = {2021},
  month = {Mar},
  publisher = {American Physical Society},
  doi = {10.1103/PhysRevB.103.094423},
  url = {https://link.aps.org/doi/10.1103/PhysRevB.103.094423}
}

\end{document}